\documentclass[journal=jacsat,manuscript=article]{achemso}

\usepackage{graphicx}
\usepackage{setspace}

\title{Learning the Hydrophilic, Hydrophobic and Aromatic Character of Amino Acids  from their
Interfacial Thermal Conductance in Water}

\author{Heydar Hamzi}
\affiliation[IUK]
{Advanced Simulation and Computing Laboratory (ASCL), Mechanical Engineering Department, Imam Khomeini International University, Qazvin, Iran.}
\author{Ali Rajabpour}  
\affiliation[IUK]
{Advanced Simulation and Computing Laboratory (ASCL), Mechanical Engineering Department, Imam Khomeini International University, Qazvin, Iran.}
\author{\'Edgar Rold\'an}
\affiliation[ICTP]
{The Abdus Salam International Center for Theoretical Physics, Strada Costiera 11, 34151 Trieste, Italy}
\author{Ali Hassanali}
\affiliation[ICTP]
{The Abdus Salam International Center for Theoretical Physics, Strada Costiera 11, 34151 Trieste, Italy}
\email{ahassana@ictp.it}

\begin{document}
\maketitle
\doublespacing

\begin{abstract}
In this study, the thermal relaxation of the 20 naturally occurring amino-acids in water is investigated using transient non-equilibrium molecular-dynamics simulations. By modeling the thermal relaxation process, the relaxation times of the amino-acids in water occurs over a timescale covering 2–5 ps. For the hydrophobic amino acids, the relaxation time is controlled by the size of the hydrocarbon side chain, while for hydrophilic amino acids, the number of hydrogen bonds do not significantly affect the timescales of the heat dissipation. Our results show that the interfacial thermal conductance at the amino-acid water interface is in the range of~40-80 MWm$^{-2}$K$^{-1}$. Hydrophobic and aromatic amino acids tend to have a lower interfacial thermal conductance. Notably, we  reveal that amino acids  can be classified, in terms of their thermal relaxation times and molar masses, into simply connected phases with the same hydrophilicity, hydrophibicity and aromaticity.
\end{abstract}

\section{Introduction}

In recent years, experimental studies and computer simulations have extensively studied heat transfer in biomolecules such as proteins and peptides. Most of the work on heat transfer in these molecules focuses on the calculation of thermal properties, including the heat capacity \cite{makhatadze1990a,privalov1990a,g1995a}, thermal diffusivity \cite{koehler2017a,yu2003a}, and the investigation of the process and path of energy exchange with the surrounding environment \cite{henry1986a,nguyen2010a,zhang2007a}, as well as obtaining interfacial thermal conductance \cite{leitner2008a,xue2003a,lervik2009a}. 

During last few decades, there have been quite a few theoretical studies investigating heat relaxation of biological systems in the gas phase \cite{henry1986a,nguyen2010a,okazaki2001a,chu1995a}. Henry et al. \cite{henry1986a}, investigated the cooling of two heme proteins using molecular dynamics simulations to predict the rate of heme vibrational energy dissipation in the protein. They found that the cooling occurs on a picosecond time scale and the decay of the vibrational temperature is non-exponential. 

There are a few works that have studied the role of the fluid surrounding the protein or peptide in order to characterize and understand how it changes the heat transfer process. The studied fluids have been sometimes hydrocarbon fluids. Sukowski et al. \cite{sukowski1990a} investigated the cooling of hot Azulene in various hydrocarbon solvents from both experimental and theoretical points of view and showed that the vibrational excess energy of Azulene is transferred to the solvent molecules by isolated binary collision. Botan et al.\cite{botan2007a} investigated energy transport through a helix in chloroform in a combined experimental-theoretical approach. They found that most of the excess heat was rapidly dissipated on a timescale of 0.5 ps into the solvent and the residual energy  propagates along the helix in a diffusive-like process through the peptide chain.

Most protein interactions occur in biological environments containing water. Besides the aforementioned studies, there have also been other works investigating heat transfer of proteins in aqueous environments \cite{lervik2010a,xu2014a,shaffer2016a,helbing2012a,lian1994a,park2009a}. The structure and physical properties of the water interface with other materials are important factors for understanding the microscopic processes in biology and the design of new nanoscale systems. Water as a polar liquid behaves differently in hydrophilic and hydrophobic interfaces \cite{zhang2002a,lum1999a,cicero2005a}. This difference is clearly observed when studying the Kapitza length \cite{kapitza1971a} and the interfacial thermal conductance \cite{swartz1989a}. For instance, Ge et al. \cite{ge2006a} by investigating the effect of polarity on the interfacial thermal conductance at hydrophilic and hydrophobic interfaces with water, showed that the Kapitza length at hydrophobic interfaces is a factor of 2-3 larger than the Kapitza length at hydrophilic interfaces.

Heat transfer within a protein occurs through its amino-acid building blocks \cite{leitner2008a,botan2007a}. There are twenty naturally occurring amino-acids which could be classified in different ways. One of the most popular chemical classifications involves the affinity of the amino acids for water namely hydrophobic, hydrophilic and amphiphilic amino-acids. There have been several measures that have been constructed to quantify the hydrophobicity of an amino acid\cite{KAPCHA2014484}. Despite many investigations that have been conducted on heat transport in different proteins and peptides, their building blocks have not been sufficiently studied in terms of their underlying heat transfer properties. Furthermore, the correlations between the chemistry of different amino acids such as their relative hydrophobic/hydrophilic character and how this changes the heat dissipation in water and in protein environments is unknown. 

In this study, by employing non-equilibrium molecular dynamics, we simulate the energy exchange process between the 20 single amino-acids with water and calculate their thermal relaxation times. We also infer the interfacial thermal conductance between all amino-acids and water. Interestingly, we find that for hydrophobic amino acids, the heat dissipation occurs on a slower time scale and is affected by the size of the hydrophobic side-chain. On the other hand, for hydrophilic amino acids, the relaxation time appears to be independent of the number of hydrogen bonds that the amino acid forms with the solvent. Aromatic amino acids are characterized by the slowest relaxation times which is likely to play an important role in their unique optical properties.

\section{Computational Details}

A transient non-equilibrium MD simulation was employed to compute the thermal relaxation times of amino-acids in water.  Figure 1 shows the chemical composition of these amino-acids and their side chains. These amino-acids are subdivided into three groups in terms of polarity of the side-chain namely the hydrophobic, hydrophilic and ampiphilic amino acids as is commonly done in biochemistry textbooks\cite{Ringe2003}.

\begin{figure}[h]
\centering
\includegraphics[scale=0.9]{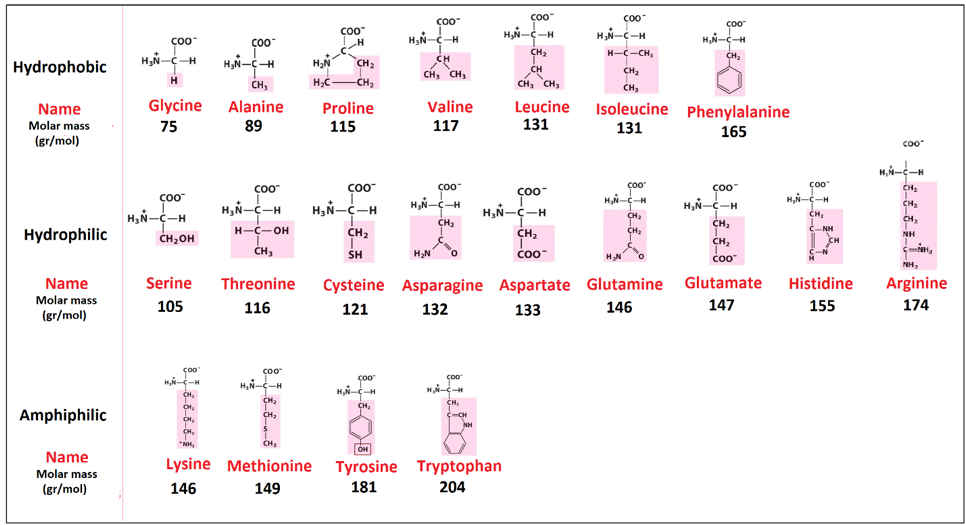}
\caption{Schematic view of the 20 naturally occurring amino-acids classified in three different groups: hydrophobic, hydrophilic and amphiphilic. Also shown are the molar masses.}
\end{figure}

In these transient molecular dynamics simulations, the amino-acid and water are thermostated to different temperatures in order to reach a non-equilibrium steady-state condition. Thereafter, one performs microcanonical simulations and using the temperature decay transient, infer the interfacial thermal conductivity. In this work, we use the lumped capacitance model to extract the interfacial thermal conductance of the naturally occurring amino acids in water. In summary, the procedure involves the following steps \cite{incropera2002a}. Firstly, the time-dependent temperature evolution during the non-equilibrium dynamics is described by the following equation:

\begin{equation}
C_p\frac{\textrm{d} T}{\textrm{d} t}=-AG[T(t)-T_f],
\end{equation}
where $C_p$ is the heat capacity of amino-acid, $A$ its surface area and G is the interfacial thermal conductance. Solving this equation leads to the following solution:

\begin{equation}
\frac{T(t)-T_f}{T_i-T_f}=\exp\left(\frac{-AG}{C_p}t\right).
\end{equation}
Defining a relaxation time as $\tau=-C_p / AG$  leads to:
\begin{equation}
\frac{T(t)-T_f}{T_i-T_f}=\exp\left(-\frac{t}{\tau}\right),
\end{equation}
where $T_i$ is the initial temperature of the amino-acid and $T_f$ is the final temperature which equilibrates to the bath temperature. Equation (3) can be fitted to the simulation data to calculate the relaxation time. The interfacial thermal conductance is then determined from the following equation:

\begin{equation}
G=\frac{C_p}{A\tau},
\end{equation}
where $A$ is the solvent accessible surface area (SASA) obtained from GROMACS \cite{eisenhaber1995a}. The heat capacity for the amino-acid is defined as the enthalpy ($H$) variations over a certain temperature change:
\begin{equation}
C_p=\left(\frac{\partial H}{\partial T}\right)_p.
\end{equation}

It should be noted that besides the lumped capacitance model, there are other theoretical models that can be used to investigate transient heat transfer in biomolecules. In a previous study, we showed that the lumped capacitance model successfully reproduces thermal transport for a silver particle in water~\cite{rajabpour2019}.

\section{Molecular Dynamics Simulations}

\subsection{Amino Acids in Water}

All simulations were performed using the GROMACS \cite{berendsen1995a,spoel2005a} package version 4.5.4. Water was modeled using the TIP4P model \cite{jorgensen1983a} and the interactions of all amino-acids described by the OPLS-AA force field \cite{jorgensen1996a,kaminski2001a}. The Leap-frog integrator was employed in MD simulations and the time step was set to 1 $fs$. Constraints were applied to all bonds with hydrogen atoms using the shake algorithm\cite{ryckaert1977a}. The structure of the every amino-acid was firstly minimized using the steepest descent method. A periodic cube box with a length of 3 $nm$ was selected and each amino acid was placed at the center of the box. A schematic setup of the simulation is shown in Figure 2.

\begin{figure}[h]
\centering
\includegraphics[scale=0.7]{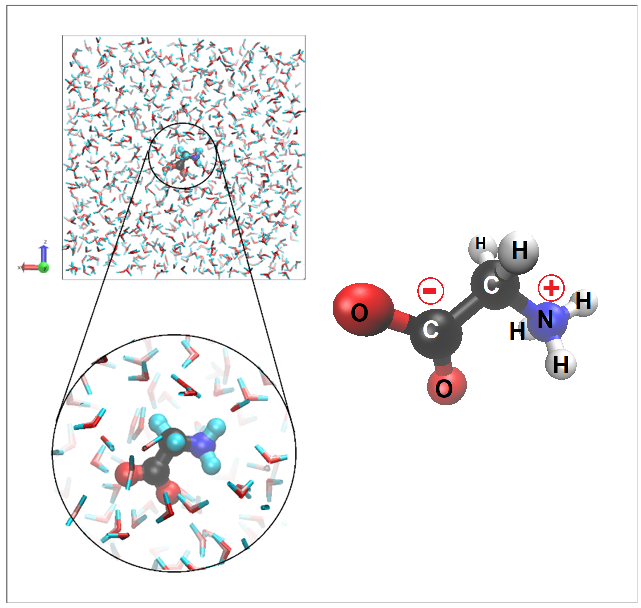}
\caption{Simulation box consisting of an amino-acid (glycine in this case in its zwitterionic state) at the center}
\end{figure}

The equilibrium  simulations were performed by first thermalizing the whole system at 300 K and 1 bar. Then, using NVT ensemble, two Nosé–Hoover thermostats were applied separately on amino-acid and water for 2.5 $ns$ to reach the target temperatures of 300 and 350 K, respectively\cite{nos1984a,nos1984b,hoover1985a}. From the last 2 ns of this data,  2000 initial conditions for the subsequent non-equilibrium simulations were chosen from configurations separated by 1 ps. In the final step, a non-equilibrium simulation using NVT ensemble was performed for 15 $ps$ in which the time constant of the amino-acid thermostat was set to infinity, but the thermostat on the solvent controlled the water temperature at 300 K with a time constant of 5 $ps$ during the energy relaxation process. The average temperature profile from all the initial conditions was determined  and fitted to Equation (3) and the relaxation time of each amino-acid was computed.

According to equation (4), the calculation of interfacial thermal conductance requires also the heat capacity of the amino acids. The heat capacity was calculated from direct numerical differentiation of the variation of internal energy of the amino-acid by varying the temperature changed from 290 K to 310 K with steps of 1 K.  These simulations were conducted in the NVT ensemble for the amino-acid in the gas-phase using a box length of 3nm.

\section{Results}

\subsection{Amino Acids in Water}

Figure 3 shows the temperature relaxation profiles obtained from the averages of 2000 trajectories for 4 amino acids: Glycine and Valine are hydrophobic, Asparagine hydrophilic and finally Tryptophan which is an aromatic amphiphilic amino acid. Interestingly, we observe some subtle differences in the relaxation times for these four different chemistries. In the inset, we also show a semi-log plot of the temperature as a function of time which shows clearly that despite the different chemical groups forming the side-chains, an exponential decay is observed giving credence to the application of the lumped-capacitance model.  Interestingly, the aromatic amino acid Tryptophan exhibits the slowest relaxation time of 5.3ps. On the other hand, glycine and asparagine which have starkly different side chains (see Figure 1), display very similar relaxation times in the heat dissipation. Valine, a hydrophobic amino acid, is characterised by relaxation times intermediate between Tryptophan and Glycine/Valine. Table 1 populates the relaxation times obtained for all the 20 natural amino acids along with other parameters used in the lumped capacitance model which will be discussed next.

\begin{figure}[h]
\centering
\includegraphics[scale=0.5]{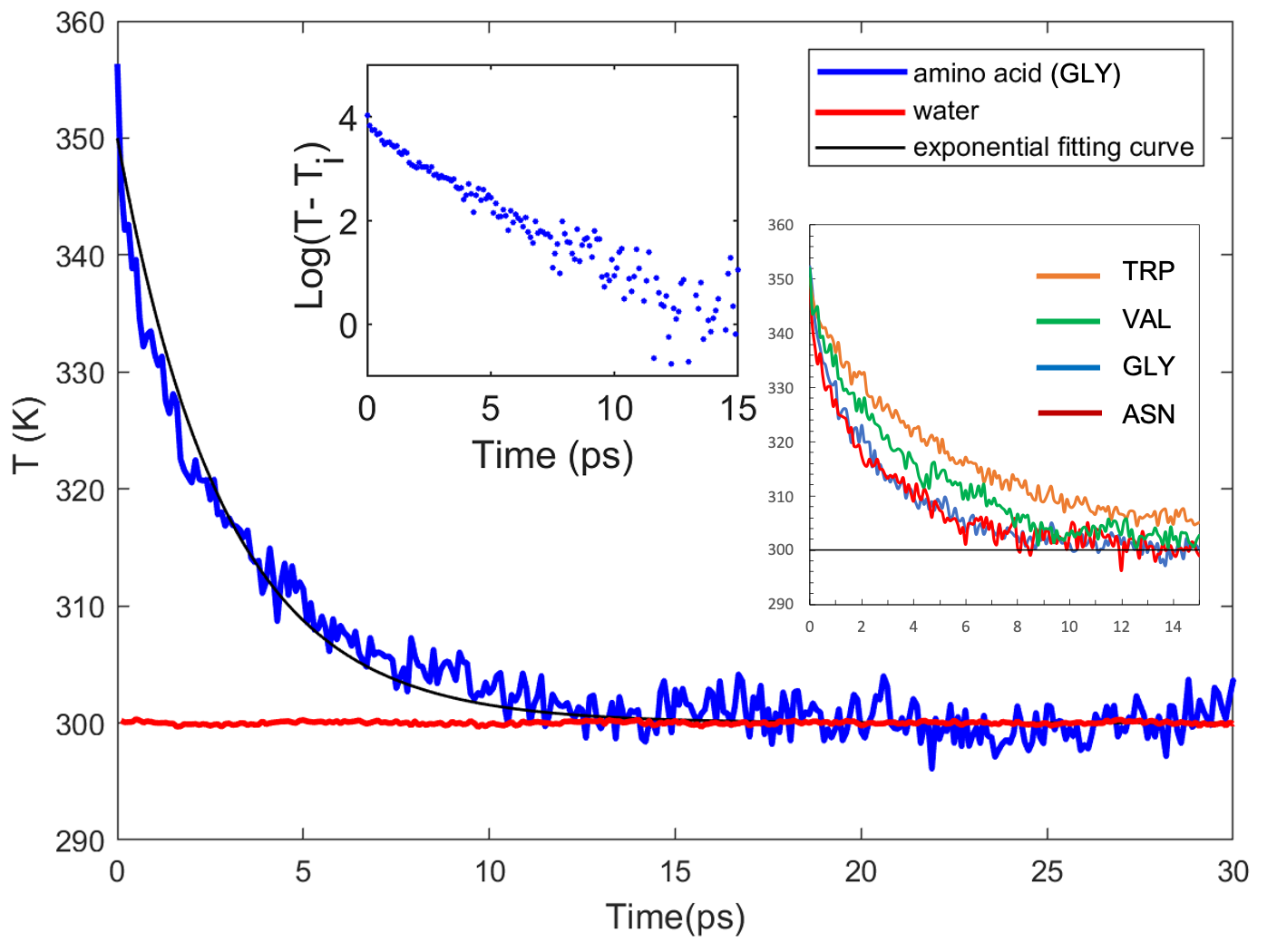}
\caption{Temperature profile associated with the relaxation and heat dissipation of glycine (blue) and temperature of water (red). The insets show the relaxation time of glycine on a semi-log scale and the temperature relaxation comparing glycine to valine, tryptophan and asparagine.}
\end{figure}

\begin{table}[htbp]
  \centering
  \caption{Calculated thermal properties of amino-acids}
    \begin{tabular}{cccccccccccccc}
    \multicolumn{2}{c}{amino  acid} & \multicolumn{1}{p{4em}}{$M$ \newline{}(gr/mol)} & \multicolumn{1}{p{4.07em}}{$C_p$ \newline{}(kJ/Kg K)} & \multicolumn{1}{p{4.21em}}{SASA\newline{} (\AA$^2$)} &       & \multicolumn{3}{p{5.36em}}{ $\tau$ \newline{} (ps)} &       & \multicolumn{3}{p{5.645em}}{$G$ \newline{}(MW/m\AA$^2$ K)} &  \\
    Hydrophobic & GLY   & 75.0  & 2.43  & 218  $\pm$5   &       & 2.34  &  $\pm$    & 0.10  && 59.30 &  $\pm$    & 2.54  &  \\
          & ALA   & 89.0  & 2.80  & 246  $\pm$2  && 2.83  &  $\pm$& 0.12  &       & 59.55 &  $\pm$    & 2.53  &  \\
          & PRO   & 115.1 & 2.88  & 279  $\pm$1  && 2.98  &  $\pm$& 0.12  &       & 66.21 &  $\pm$    & 2.72  &  \\
          & VAL   & 117.1 & 3.02  & 294  $\pm$4  && 3.58  &  $\pm$& 0.11  &       & 55.87 &  $\pm$    & 1.64  &  \\
          & LEU   & 131.1 & 3.32  & 322  $\pm$6  && 3.80  &  $\pm$& 0.09  &       & 59.07 &  $\pm$    & 1.40  &  \\
          & ILE   & 131.1 & 3.26  & 318  $\pm$5  && 4.15  &  $\pm$& 0.09  &       & 53.82 &  $\pm$    & 1.16  &  \\
          & PHE   & 165.1 & 2.79  & 358  $\pm$3  && 5.33  &  $\pm$& 0.16  &       & 40.06 &  $\pm$    & 1.17  &  \\
    Hydrophilic & SER   & 105.0 & 2.66  & 254  $\pm$4  &       & 2.61  &  $\pm$    & 0.11  && 69.88 &  $\pm$    & 2.97  &  \\
          & THR   & 119.0 & 2.92  & 275  $\pm$6  & & 2.74  &  $\pm$& 0.09  & & 76.72 &  $\pm$& 2.50  &  \\
          & CYS   & 121.0 & 2.21  & 269  $\pm$4  & & 3.09  &  $\pm$& 0.12  & & 53.35 &  $\pm$& 2.02  &  \\
          & ASN   & 132.0 & 2.55  & 295  $\pm$7  & & 2.35  &  $\pm$& 0.08  & & 80.63 &  $\pm$& 2.68  &  \\
          & ASP   & 133.1 & 2.53  & 284  $\pm$3  & & 3.10  &  $\pm$& 0.10  & & 63.44 &  $\pm$& 2.07  &  \\
          & GLN   & 146.1 & 2.99  & 326  $\pm$2  & & 2.76  &  $\pm$& 0.07  & & 80.68 &  $\pm$& 2.05  &  \\
          & GLU   & 147.1 & 2.52  & 307  $\pm$4  & & 3.20  &  $\pm$& 0.07  & & 62.66 &  $\pm$& 1.37  &  \\
          & HIS   & 155.1 & 2.71  & 328  $\pm$4  & & 3.37  &  $\pm$& 0.10  & & 63.18 &  $\pm$& 1.78  &  \\
          & ARG   & 174.1 & 3.00  & 395  $\pm$7  & & 3.34  &  $\pm$& 0.06  & & 65.69 &  $\pm$& 1.18  &  \\ Amphiphilic & LYS   & 146.1 & 2.99  & 340  $\pm$4  &       & 4.51  &  $\pm$    & 0.07  && 47.33 &  $\pm$& 0.74  &  \\
          & MET   & 149.0 & 2.70  & 327  $\pm$5 & & 4.10  &  $\pm$& 0.07  & & 49.89 &  $\pm$& 0.85  &  \\
          & TYR   & 181.1 & 2.78  & 371  $\pm$6 & & 5.01  &  $\pm$& 0.11  & & 44.95 &  $\pm$& 1.02  &  \\
          & TRP   & 204.1 & 2.66  & 395  $\pm$5 & & 5.31  &  $\pm$& 0.14  & & 42.96 &  $\pm$& 1.12  &  \\
    \end{tabular}%
  \label{tab:addlabel}%
\end{table}%

In order to understand better the correlations between the relaxation times and the chemistry of the amino acid chains, Figure 4 shows a scatter plot of the relaxation times and molar mass for all the amino acids. In the left panel of Figure 4, the data is colored based on the classification of hydrophobic (colored blue), hydrophilic (colored red) and ampiphilic (colored green). In the right panel however, the amino acids are color coded based on being aromatic or not.  

Overall, we observe that for both the hydrophobic and ampiphilic amino acids, the relaxation times seem to we well correlated with the molar mass. On the other hand for the hydrophilic amino acids, the variations in the relaxation times are much less pronounced.  Figure 4 also clusters the data colored based on the the hydrophobic, hydrophilic and ampiphilic amino acids. For the hydrophobic amino acids, it appears as though that the relaxation times are controlled essentially by the number of atoms forming the side chain. For the hydrophilic amino acids, it is interesting to note that despite the very different chemistries of the side chains, the relaxation times do not change so drastically. In particular, we see for example, that both Aspartic and Glutamic acid which are negatively charged have very similar relaxation times to Histidine and Arginine the latter of which is positively charged. Cysteine which has a sulfur atom is also characterized by similar relaxation times. 

Comparing Asparagine to Aspartic acid and Glutamine to Glutamic acid, we observe the relative role of both the negative charge and the size of the side chain. The acids tend to have a slightly longer relaxation times owing to the fact that the negative charge polarizes the water more and therefore the heat dissipation from the modes of the amino acid into the solvent takes slightly longer.

In the right panel of Figure 4, we color-coded the plots into the aromatic and non-aromatic amino acids. We observe clearly here that all the aromatic amino acids are characterized by longer relaxation times and essentially form a separate cluster compared to the non-aromatics. The slower timescale associated with the aromatics is attributed to their planar rigid structure and therefore these amino acids are able to trap heat for longer times before it is dissipated into the surrounding solvent. These longer relaxation times are likely to play a critical role in their enhanced fluorescence\cite{Lakowicz2006}. Furthermore the relaxation time and the mass allow us to identify the different types of amino acids within simply-connected clusters. With the growth and use of data-science based approaches in physical chemistry\cite{ansari2019,laiochemrev2021}, properties such as the statistics associated with the heat dissipation times related to thermal properties of bio-materials like proteins and DNA, may be an interesting frontier to explore in the future.

\begin{figure}[h]
\centering
\includegraphics[scale=0.5]{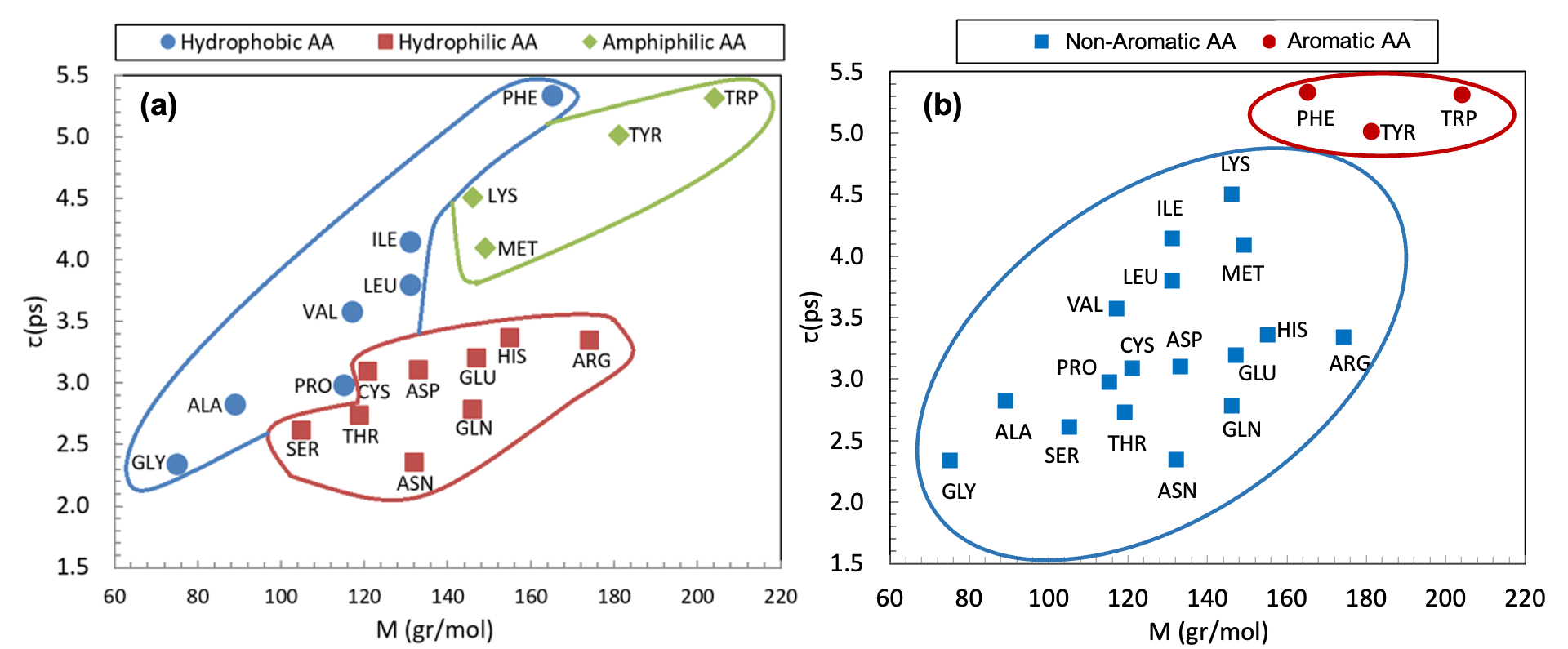}
\caption{Correlations obtained between the thermal relaxation times and molar masses of all the amino acids. In panel a the data is partitioned with hydrophobic, hydrophilic and amphiphilic amino acids, while in panel b the the partitioning is done with respect to the aromatic and non-aromatic amino acids.}
\end{figure}


By computing both the specific heat capacity of each amino acid and its solvent accessible surface area, we inferred the interfacial thermal conductance using Equation 4. The interfacial thermal conductance for some water-immersed proteins has been reported at the order of 100-270 MW/m$^2$.K \cite{lervik2010a}. In our work, the thermal conductance at amino-acids/water interface is expected to be lower than that reported for proteins. The left panel of Figure 5 shows the interfacial thermal conductance for the hydrophobic, hydrophilic and ampiphilic amino acids, while the right panel shows the partitioning obtained between the aromatic and non-aromatic amino acids. We observe that the interfacial thermal conductance for amino acids in water cover a range between 30-85 MW/m$^2$.K which is smaller than that of proteins. 

\begin{figure}[h]
\centering
\includegraphics[scale=0.5]{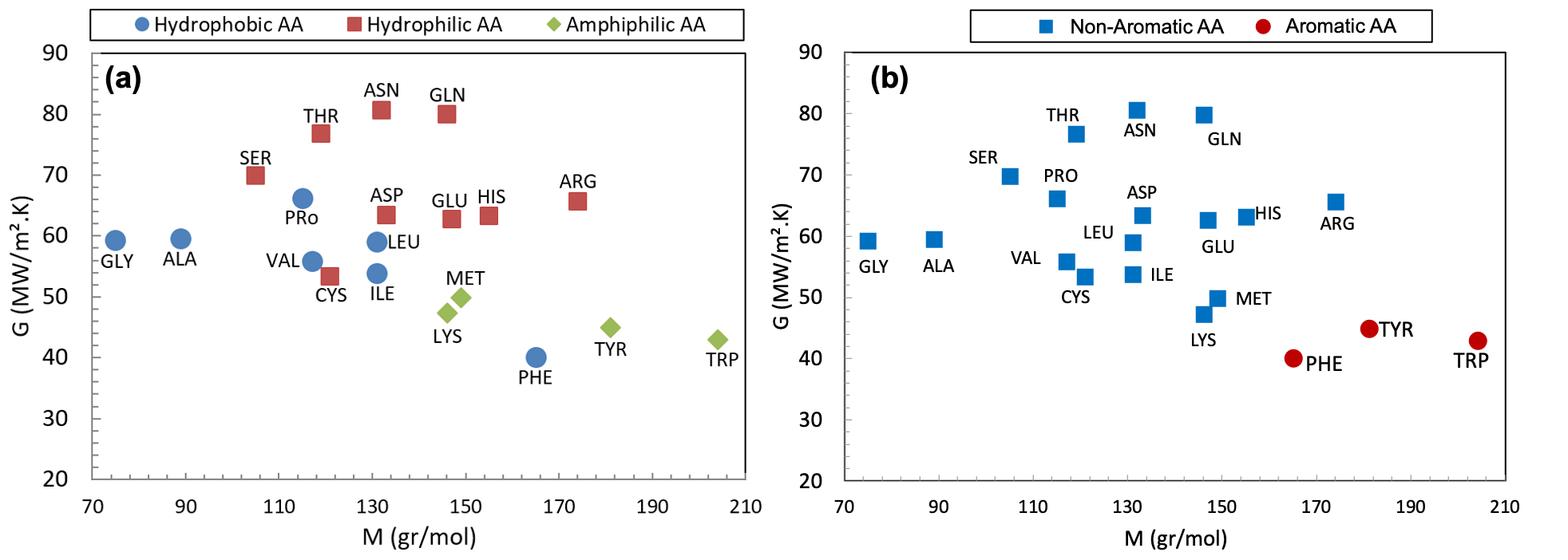}
\caption{Interfacial thermal conductance for single amino acids in water. Panel a shows the thermal conductance classified based on the hydrophobic, hydrophilic and amphiphilic amino acids, while in panel b the partitioning is done with respect to the aromatic and non-aromatic amino acids.}
\end{figure}

Consistent with what we observe in Figure 4, hydrophilic amino acids have, on average a higher interfacial thermal conductance followed by the hydrophobic ones and then the amphililic amino acids. The stronger the interactions between a particle and the surrounding solvent, the larger the thermal conductance\cite{xue2003a,wang2011a,shenogina2009a}. Hence, the hydrophilic amino acids which are characterized by stronger interactions with the surrounding solvent have a higher interfacial thermal conductance. The right panel of Figure 5 also shows the thermal conductance for aromatic and non-aromatic amino acids - the aromatics are again clearly endowed by a smaller interfacial thermal conductance which may play an important role in making them the most optically active chromophores in nature\cite{Lakowicz2006}. Table. 1 shows all the data that was used to construct the plots shown in Figure 4 and Figure 5 including the solvent-accessible surface area, the relaxation time $\tau$ and the interfacial thermal conductance for each amino acid.


\begin{figure}[h]
\centering
\includegraphics[scale=0.5]{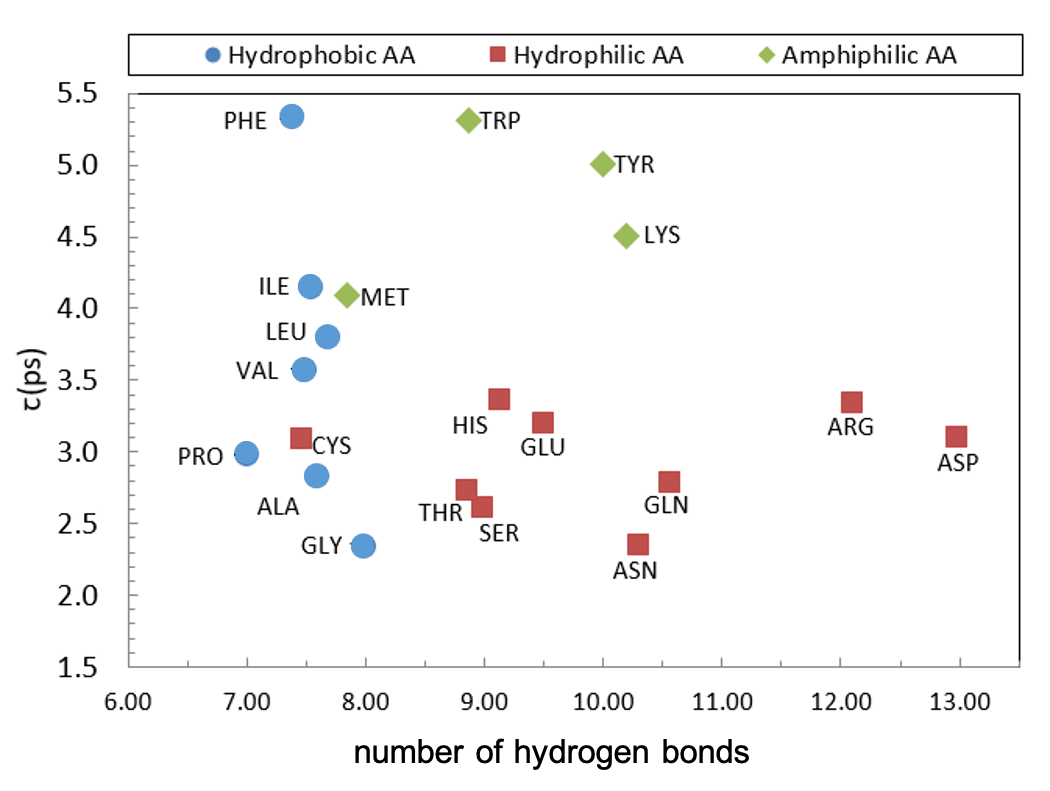}
\caption{Scatter plot between thermal relaxation times and the total number of hydrogen bonds (donating and accepting) with the environment.}
\end{figure}

One of the distinguishing features of the different amino acids is their ability to form hydrogen bonds with the surrounding water molecules. In this regard, for the hydrophobic amino acids, hydrogen bonds can only be formed between the N and C termini and the water, while for the hydrophilic and ampiphilic amino acids, the side chains can also donate or accept hydrogen bonds. Figure 6 illustrates a scatter plot showing the behavior between $\tau$ and the average number of hydrogen bonds that the amino acid makes with the surrounding water. The number of hydrogen bonds for the hydrophobic amino acids ranges between 7-8 which arises from the N and C termini donating and accepting $\sim$3 to 4 hydrogen bonds each. Thus, for the hydrophobic amino acids, the time scale of heat dissipation is modulated almost entirely by the size of the hydrophobic group. 

The hydrophilic amino acids on the other hand, seem to display rather different behavior from the hydrophobic ones. Specifically, the thermal relaxation times for the hydrophilic amino acids is independent of the number of hydrogen bonds. Both Aspartic acid and Arginine which have a large number of hydrogen bonds arising from their side chains,  have very similar relaxation times to cysteine which forms a similar number of hydrogen bonds as the hydrophobic amino acid alanine. As seen in Figure 4, the overall mass of the amino acid appears to be a better predictor for the timescales of heat dissipation.

\section{Discussion and Conclusion}

In this work, we have used molecular dynamics simulations in combination with a simple heat diffusion model to characterize the heat dissipation mechanism of amino acids in water. We use this subsequently, to infer the interfacial thermal conductance of the 20 naturally occurring amino acids in liquid water. Heat dissipation, as inferred from this model, is strongly coupled to the polarity of the amino acids. 

We have found that the relaxation times of the amino-acids in water at 300 K, is in the range of $2–5$ps. For the hydrophobic amino acids, the critical factor controlling the relaxation time are the hydrophobic side chains. On the other hand, for the hydrophilic amino acids, their ability to form hydrogen bonds with the solvent does not appear to change the heat dissipation time. 

Using the lumped capacitance model, we infer an interfacial thermal conductance at the amino acid-water interfaces to be on the order of 40-80 MWm$^{-2}$K$^{-1}$ which is significantly lower than the value reported for proteins. The interfacial thermal conductance of hydrophilic amino-acids was found more than that of hydrophobic amino-acids which is related to the stronger interactions that they form with with the surrounding solvent. Interestingly, the aromatic amino acids tend to be characterized by slower thermal relaxation times and lower interfacial thermal conductance. Furthermore, while in this work we have focused exclusively on the average temperature relaxation, the higher order moments of the temperature may also cary important information. It will be interesting in the future to analyze the non-equilibrium temperature fluctuations of amino acids from the viewpoint of stochastic thermodynamics~\cite{hatano2001steady,karimi2020quantum}.

\providecommand{\latin}[1]{#1}
\makeatletter
\providecommand{\doi}
  {\begingroup\let\do\@makeother\dospecials
  \catcode`\{=1 \catcode`\}=2 \doi@aux}
\providecommand{\doi@aux}[1]{\endgroup\texttt{#1}}
\makeatother
\providecommand*\mcitethebibliography{\thebibliography}
\csname @ifundefined\endcsname{endmcitethebibliography}
  {\let\endmcitethebibliography\endthebibliography}{}

\end{document}